\documentclass[fleqn,twoside]{article}
\usepackage{espcrc2,graphicx}

\newcommand{\ind}[2]{^{#1}_{\mbox{\scriptsize #2}}}
\newcommand{\al}[2]{\alpha\ind{#1}{#2}}
\newcommand{\tal}[2]{\widetilde{\alpha}\ind{#1}{#2}}
\newcommand{\hal}[2]{\widehat{\alpha}\ind{#1}{#2}}
\newcommand{\bt}[2]{\beta\ind{#1}{#2}}
\newcommand{\ro}[1]{\rho^{(#1)}}

\def\DCI{D^{\mbox{\tiny CI}}}
\def\LQCD{$\Lambda_{\mbox{\tiny QCD}}$ }
\def\sfa{\mbox{\sf A}}
\def\nf{n_{\mbox{\scriptsize f}}}
\def\KL{K\"all\'en--Lehmann }

\title{Nonperturbative aspects of
the QCD analytic invariant charge}

\author{A.V. Nesterenko
\address{Centre de Physique Theorique de l'\'Ecole Polytechnique, 
91128 Palaiseau Cedex, France \\ 
Unit\'e Mixte de Recherche du CNRS (UMR 7644)}\address{Bogoliubov 
Laboratory of Theoretical Physics, JINR, 
Dubna 141980, Russian Federation}\thanks{Partially supported by 
RFBR (grants 02-01-00601 and 00-15-96691).}
\addtocounter{footnote}{-1}}
       
\begin{document}

\begin{abstract}
In the framework of the analytic approach to Quantum Chromodynamics a
new model for the strong running coupling has recently been
developed. Its underlying idea is to impose the analyticity
requirement on the perturbative expansion of the renormalization
group $\beta$~function for restoring the correct analytic properties
of the latter. The proposed model possesses a number of appealing
features. Namely, the analytic invariant charge has no unphysical
singularities at any loop level; it contains no free parameters; it
has universal behavior both in ultraviolet and infrared regions at
any loop level; and it possesses a fair higher loop and scheme
stability. The extension of this model to the timelike region
revealed the asymmetrical behavior of the running coupling in the
intermediate- and low-energy domains of spacelike and timelike
regions, that is essential when one handles the experimental data.
The developed approach enables one to describe, in a consistent way,
various strong interaction processes both of perturbative and
intrinsically nonperturbative nature.
\vspace{1pc}
\end{abstract}

\maketitle

\section{INTRODUCTION}

     Theoretical analysis of hadron dynamics in main part relies upon
the renormalization group (RG) method. Meantime, the construction of
exact solutions to the RG equations is still far from being feasible.
Usually, in order to describe the strong interaction processes in
asymptotical ultraviolet (UV) region, one applies the RG method
together with perturbative calculations. In this case, the outcoming
approximate solutions to the RG equations are used for quantitative
analysis of experimental data. However, such solutions have
unphysical singularities in the infrared (IR) domain, that
contradicts the basic principles of local Quantum Field 
Theory~(QFT).

     An effective way to overcome such difficulties consists in
invoking into consideration the analyticity requirement, which
follows from the general principles of local~QFT. This prescription
became the underlying idea of the so-called analytic approach to QFT,
which was first formulated in late 1950s~\cite{RedBLS}. Lately this
approach has been extended to Quantum Chromodynamics
(QCD)~\cite{ShSol}, and applied to the ``analytization'' of
perturbative series for the QCD observables~\cite{SolSh}. The term
``analytization'' means the restoring of the correct analytic
properties in the $q^2$ variable of a quantity under consideration by
making use of the \KL integral representation
\begin{equation}
\Bigl\{\,\sfa(q^2)\,\Bigr\}_{\!\mbox{\scriptsize an}} =
\int_{0}^{\infty}\! \frac{\varrho(\sigma)}{\sigma+q^2}\, 
d \sigma
\end{equation}
with the spectral function defined by the initial (perturbative) 
expression for a quantity in hand:
\begin{equation}
\varrho(\sigma) = \frac{1}{2 \pi i} \lim_{\varepsilon \to 0_{+}}
\Bigl[\sfa(-\sigma-i\varepsilon)-\sfa(-\sigma+i\varepsilon)\Bigr].
\end{equation}

\section{ANALYTIC INVARIANT CHARGE}

     First of all, let us consider the renormalization group equation
for the strong running coupling:
\begin{equation}
\label{RGGen}
\frac{d\,\ln \bigl[g^2(\mu)\bigr]}{d\,\ln\mu^2} =
\beta\Bigl(g(\mu)\Bigr).
\end{equation}
In the framework of perturbative approach the $\beta$~function on the
right-hand side of this equation  can be represented as a power
series
\begin{equation}
\label{BetaPertSer}
\beta\Bigl(g(\mu)\Bigr)\!=\! -\!\left\{
\beta_{0}\!\left[\frac{g^2(\mu)}{16 \pi^2}\right]\!+\!
\beta_{1}\!\left[\frac{g^2(\mu)}{16 \pi^2}\right]^2\!\!+...\!\right\}\!,
\end{equation}
where $\beta_{0} = 11 - 2\,\nf\,/\,3,\,$ $\beta_{1}=102 - 38 \,\nf
\,/\,3$, and $\nf$ is the number of active quarks. Introducing the
standard notations  $\al{}{s}(\mu^2)= g^2(\mu)/(4\pi)$ and
$\tal{}{}(\mu^2) = \alpha(\mu^2)\, \beta_{0}/(4\pi)$, one can rewrite
the RG equation~(\ref{RGGen}) at the $\ell$-loop level in the form
\begin{equation}
\frac{d\,\ln\bigl[\tal{(\ell)}{s}(\mu^2)\bigr]}{d\,\ln \mu^2} = -
\sum_{j=0}^{\ell-1} B_{j} \left[\tal{(\ell)}{s}(\mu^2)\right]^{j+1},
\end{equation}
where $B_{j} = \beta_{j}/\beta_{0}^{j+1}$. It is well-known that the
solution to this equation has unphysical singularities at any loop
level. Thus, the strong running coupling  gains unphysical
singularities due to perturbative approximation of the
renormalization group  $\beta$~function~(\ref{BetaPertSer}). However,
the fundamental principles of local QFT require the invariant charge
$\al{}{s}(q^2)$ to have a definite analytic properties in the $q^2$ 
variable. Namely, there must be the only cut along  the
negative\footnote{A metric with the signature $(-1,1,1,1)$ is used,
so that positive $q^2$ corresponds to a spacelike momentum transfer.}
semiaxis of real~$q^2$~\cite{ShSol}.

     In the framework of developed model~\cite{PRD1,PRD2} the
analyticity requirement is imposed on the $\beta$~function
perturbative expansion for restoring its correct analytic properties.
This leads to the following RG equation for the analytic invariant
charge:
\begin{equation}
\label{AnRGEq}
\frac{d\,\ln\bigl[\tal{(\ell)}{an}(\mu^2)\bigr]}{d\,\ln \mu^2} = 
- \left\{\sum_{j=0}^{\ell-1} B_{j}\Bigl[\tal{(\ell)}{s}(\mu^2)
\Bigr]^{j+1}\right\}_{\!\mbox{\scriptsize an}}\!\!.
\end{equation}
At the one-loop level this equation can be integrated explicitly
with the result~\cite{PRD1}
\begin{equation}
\label{AICOneLoop}
\al{(1)}{an}(q^2) = \frac{4\pi}{\beta_0}\,\frac{z-1}{z\,\ln z}, 
\qquad z=\frac{q^2}{\Lambda^2}.
\end{equation}
At the higher loop levels only the integral representation for the
analytic invariant charge (AIC) was derived. So, at the $\ell$-loop
level the solution to the RG equation~(\ref{AnRGEq}) takes the
form~\cite{PRD2,MPLA2}:
\begin{equation}
\label{AICHighLoop}
\al{(\ell)}{an}(q^2) = \frac{4\pi}{\beta_{0}} \, \frac{z-1}{z\,\ln z}
\,\exp\!\left[\int_{0}^{\infty}\! 
{\cal P}^{(\ell)}(\sigma)\, \frac{d \sigma}{\sigma}\right],
\end{equation}
with ${\cal P}^{(\ell)}(\sigma)\!\!=\!\!
\left[{\cal R}^{(\ell)}(\sigma) - {\cal R}^{(1)}(\sigma)\right]
\ln (1\!+\!\sigma/z)$, and
\begin{eqnarray}
{\cal R}^{(\ell)}(\sigma) &=& \lim_{\varepsilon \to 0_{+}}
\sum_{j=0}^{\ell-1} \frac{B_j}{2 \pi i}
\left\{\Bigl[\tal{(\ell)}{s}(-\sigma-i\varepsilon)\Bigr]^{j+1}
\right. \nonumber \nopagebreak \\ && \left.
-\Bigl[\tal{(\ell)}{s}(-\sigma+i\varepsilon)\Bigr]^{j+1}\right\}.
\end{eqnarray}
It is worth mentioning also that the representation of the \KL type
holds for the analytic invariant charge~(\ref{AICHighLoop})
\begin{equation}
\label{AICKL}
\al{(\ell)}{an}(q^2) = \frac{4\pi}{\beta_{0}}\int_{0}^{\infty}
\frac{\ro{\ell}(\sigma)}{\sigma+z}\,d\sigma,
\end{equation}
where the $\ell$-loop spectral density $\ro{\ell}(\sigma)$ is
defined by Eqs.~(18)--(19) of Ref.~\cite{PRD2}.

     The analytic invariant charge~(\ref{AICHighLoop}) possesses a
number of appealing features. First of all, it has no unphysical
singularities at any loop level. Then, the developed model contains
no adjustable parameters, i.e., similarly to perturbative approach,
\LQCD remains the basic characterizing parameter of the theory. The
AIC also incorporates UV asymptotic freedom with IR enhancement in a
single expression, that plays an essential role in applications of
developed model.

\begin{figure}[t]
\noindent
\includegraphics[width=60mm]{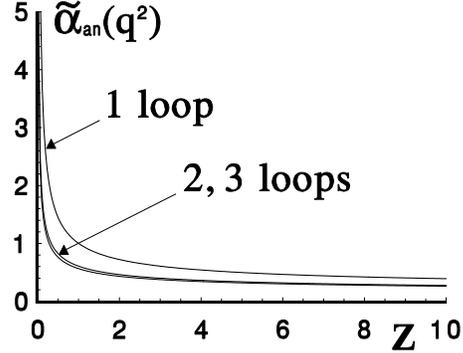}
\caption{The QCD analytic invariant charge~(\protect\ref{AICHighLoop})
at different loop levels, $z=q^2/\Lambda^2$.}
\label{Fig:AIC}
\end{figure}

\begin{figure}[t]
\noindent
\includegraphics[width=59.5mm]{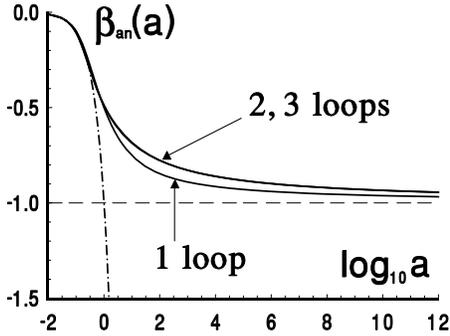}
\caption{The $\beta$ function corresponding to the
AIC~(\protect\ref{AICHighLoop}) at different loop levels. The
one-loop perturbative result is shown as the dot-dashed curve.}
\label{Fig:BetaHL}
\end{figure}
     
     Figure~\ref{Fig:AIC} shows the running coupling $\tal{}{an}(q^2)
= \al{}{an}(q^2)\,\beta_{0}/(4 \pi)$ at different loop levels. In
particular, it is follows from this figure that
AIC~(\ref{AICHighLoop}) possesses a good higher loop stability.  The
study of the relevant $\beta$~function at the higher loop levels 
also enables one to elucidate the asymptotic behavior of the AIC. So,
the $\beta$~function in hand
\begin{equation}
\label{BetaHLDef}
\bt{(\ell)}{an}(a) = \frac{d \, \ln a(\mu^2)}{d \, \ln \mu^2}, \qquad
a(\mu^2) \equiv \tal{(\ell)}{an}(\mu^2)
\end{equation}
was proved to have the universal asymptotics at any loop
level~\cite{MPLA2}. Namely, it coincides with the perturbative result
$\bt{(1)}{s}(a)=-a$ at small values of running coupling and tends to
the minus unity at large~$a$ (see Figure~\ref{Fig:BetaHL}). In turn,
this trait of the  $\beta$~function implies that the
AIC~(\ref{AICHighLoop}) itself possesses  the universal asymptotics
both in the ultraviolet $[\tal{(\ell)}{an}(q^2) \simeq
1/\ln(q^2/\Lambda^2)]$ and infrared $[\tal{(\ell)}{an}(q^2) \simeq
\Lambda^2/q^2]$ regions at any loop level. The detailed analysis of
the properties of analytic running coupling can be found in
Refs.~\cite{MPLA2,MPLA1}.

\section{TIMELIKE REGION}

     For the consistent description of some strong interaction
processes one has to employ the running coupling in the timelike
region \mbox{($s = -q^2 >0$)}.  By making use of the prescription
elaborated in Ref.~\cite{MiltSol} the continuation of AIC
to the timelike domain has recently been performed
\begin{equation}
\label{AICTL}
\hal{(\ell)}{an}(s) = \frac{4 \pi}{\beta_0} \,
\int_{s/\Lambda^2}^{\infty}
\ro{\ell}(\sigma) \, \frac{d \sigma}{\sigma}
\end{equation}
(see Ref.~\cite{PRD2} for the details). Here the running coupling in
the spacelike region is denoted by $\alpha(q^2)$ and in the timelike
region by~$\widehat{\alpha}(s)$. It is worth noting also that the
obtained result~(\ref{AICTL}) confirms the hypothesis due to
Schwinger concerning proportionality between the $\beta$ function and
relevant spectral density (see Refs.~\cite{PRD2,MiltSol}). 

\begin{figure}[t]
\noindent
\includegraphics[width=60mm]{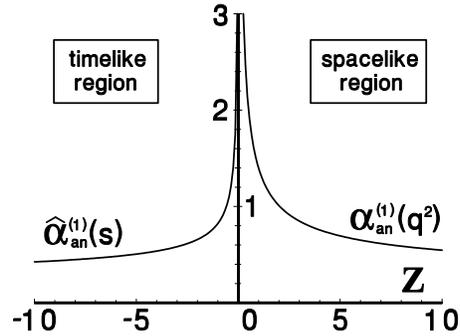}
\caption{The one-loop analytic running coupling in the 
spacelike~(\protect\ref{AICOneLoop}) and
timelike~(\protect\ref{AICTL}) regions.}
\label{Fig:AICST}
\end{figure}

     The plots of the one-loop analytic invariant charge in the
spacelike and timelike domains are shown in Figure~\ref{Fig:AICST}.
In the ultraviolet limit these functions have identical behavior
determined by the asymptotic freedom. However, there is asymmetry
between them in the intermediate- and low-energy regions. The
relative difference  between $\hal{(1)}{an}(s)$ and
$\al{(1)}{an}(q^2)$ is about several percents at the scale of the
$Z$~boson mass, and increases when approaching the infrared domain.
Apparently, this circumstance must be taken into account when one
handles the experimental data.

\section{PHENOMENOLOGICAL\\ APPLICATIONS}
\label{Sec:Applic}

     A decisive test of the self-consistency of any model for the
strong interaction is its applicability to description of diverse QCD
processes. 

\begin{figure}[t]
\noindent
\includegraphics[width=60mm]{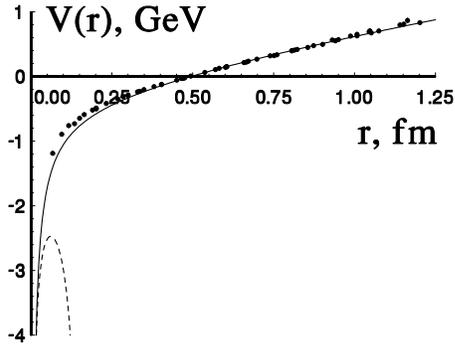}
\caption{The constructed quark--antiquark potential $V(r)$ (solid
curve) and the quenched lattice data ($\protect\bullet$);
$\Lambda=670\,$MeV, $\nf=0$. The dashed curve denotes the  relevant
perturbative result.}
\label{Fig:Vr}
\end{figure}

     It was shown \cite{PRD1,Potent} that the analytic running
coupling~(\ref{AICOneLoop}) explicitly leads to the confining  static
quark--antiquark potential
\begin{equation}
\label{VrConf}
V(r) \simeq \frac{8\pi}{3\beta_0}\frac{1}{r_0}\,
\frac{r/r_0}{2\,\ln(r/r_0)}, \qquad r \to \infty,
\end{equation}
where $r_{0}^{-1}=\Lambda\,e^C$ is a reference scale, and
$C\simeq0.577...$ denotes the Euler's constant. At the same time,
this potential has the standard behavior, determined by the
asymptotic freedom, at small distances. The potential derived is
found to be in a good agreement with the relevant lattice simulation
data~\cite{Bali} (see Figure~\ref{Fig:Vr}). The obtained one-loop
estimation  of the scale parameter $\Lambda=(670 \pm 8)\,$MeV for
$\nf=0$ active quarks corresponds to the value $\Lambda=(590 \pm
10)\,$MeV in the region of three active quarks (see also
Ref.~\cite{Potent}).

     It was found~\cite{MPLA2} that there are symmetries, which
relate the analytic invariant charge and the corresponding 
$\beta$~function in the UV and IR domains. One might anticipate that
they could also be revealed in some intrinsically nonperturbative 
strong interaction processes. It is worth mentioning here the 
recently discovered symmetry related to the size distribution of 
instantons~$D(\rho)$. Namely, the lattice simulation
data~\cite{UKQCD} for the quantity $\DCI(\rho) = D(\rho)
\left(\rho/\rho_0\right)^{N_c/6}$, where
\begin{equation}
D(\rho) = b \left[\frac{2 \pi}{\alpha(\rho)}\right]^{2 N_c}
\exp\!\left[- \frac{2 \pi}{\alpha(\rho)}\right],
\end{equation}
have been observed to satisfy the conformal inversion symmetry
$\DCI(\rho) = \DCI(\rho_0^2/\rho)$~\cite{Schrempp01}. In turn, this
relation imposes a certain constraint onto the properties of the QCD
running coupling $\alpha(\rho)$ at small and large distances.
Proceeding from this, the analytic invariant
charge~(\ref{AICOneLoop}) has recently been rediscovered and
proved~\cite{Schrempp01} to reproduce explicitly the conformal
inversion symmetry of $\DCI(\rho)$.

     The AIC has also been applied to description of the gluon
condensate ($\Lambda=(631 \pm 79)\,$MeV), inclusive  $\tau$~lepton
decay ($\Lambda=(524 \pm 68)\,$MeV), and $e^{+}e^{-}$~annihilation
into hadrons ($\Lambda=(490 \pm 76)\,$MeV).  Thus, the applications
of the model developed to description of diverse strong interaction
processes ultimately lead to congruous estimation of the
parameter~$\Lambda_{\mbox{\tiny QCD}}$. Namely, at the one-loop level
with three active quarks $\Lambda = (557 \pm 36)\,$MeV. Apparently,
this testifies that the AIC substantially incorporates, in a
consistent way, both perturbative and intrinsically nonperturbative
aspects of Quantum Chromodynamics.

\section{CONCLUSIONS}

     We have shown that the proposed way of involving the analyticity
requirement into the RG~method leads to qualitatively new features of
the strong running coupling. Ultimately, this enables one to
describe, in a consistent way, various strong interaction processes
both of perturbative and intrinsically nonperturbative nature.

\end{document}